# Robust, frequency-stable and accurate mid-IR laser spectrometer based on frequency comb metrology of quantum cascade lasers up-converted in orientation-patterned GaAs

Michael G. Hansen,[1,*] Ingo Ernsting,[1] Sergey V. Vasilyev,[1] Arnaud Grisard,[2] Eric Lallier,[2] Bruno Gérard,[3] and Stephan Schiller[1]

[1]*Heinrich-Heine-Universität Düsseldorf, Institut für Experimentalphysik, Universitätsstr. 1, 40225 Düsseldorf*
[2]*Thales Research and Technology, Campus Polytechnique, 91767 Palaiseau cedex, France*
[3]*III-V Lab, Campus Polytechnique, 91767 Palaiseau cedex, France*
[*]*michael.hansen@uni-duesseldorf.de*

**Abstract**

We demonstrate a robust and simple method for measurement, stabilization and tuning of the frequency of cw mid-infrared (MIR) lasers, in particular of quantum cascade lasers. The proof of principle is performed with a quantum cascade laser at 5.4 µm, which is upconverted to 1.2 µm by sum-frequency generation in orientation-patterned GaAs with the output of a standard high-power cw 1.5 µm fiber laser. Both the 1.2 µm and the 1.5 µm waves are measured by a standard Er:fiber frequency comb. Frequency measurement at the 100 kHz-level, stabilization to sub-10 kHz level, controlled frequency tuning and long-term stability are demonstrated.

## 1. Introduction

The MIR spectral range ($\lambda > 4.5$ µm) is of interest in both applied and fundamental spectroscopy, for diverse applications such as trace gas detection [1], cold molecules manipulation [2] and molecular frequency metrology [3, 4]. Continuous-wave (cw) laser sources in this spectral range suitable for precision experiments requiring absolute frequency knowledge [5-10] are typically complex and of limited performance (line emission only, or low power).

One approach currently pursued for enabling MIR spectroscopy with accurate frequency control is based on generating the desired radiation by down-conversion, using frequency combs [11-20] or cw near-IR (NIR) sources [10, 21, 22]. For interrogation of single molecular lines, the comb down-conversion approach is not ideal, since an individual comb line has only ~ 10 nW power.

Another approach is based on using quantum cascade lasers (QCL) as the source of spectroscopic radiation. The introduction of QCLs has revolutionized the field of MIR spectroscopy. The combination of QCLs with a tool for measurement and control of their absolute frequency is particularly interesting because modern QCLs provide all features desirable from a flexible cw spectroscopy source: high power (enabling nonlinear spectroscopy, such as saturation spectroscopy or two-photon spectroscopy), wide tuning range and good mode quality (with proper control, e.g. in combination with the external cavity technique), room-temperature operation, compact size, low power consumption.

The application of QCLs to high-resolution molecular spectroscopy and the understanding of their intrinsic frequency noise and frequency control capability has already been advanced substantially [23-26]. While linewidths of free-running room-temperature QCLs can be large (hundreds of kHz to a few MHz on the ms timescale, depending on the noise level of the current source [27]), it has been shown that they can be reduced enormously, to the Hz-level, by fast frequency control [28, 29]. Even phase-locking of QCLs has been achieved [30]. These studies indicate that QCLs are suitable for highest-resolution (Doppler-free) spectroscopy applications. However, in actual practice, frequency stabilization has so far been implemented by locking to molecular absorption lines, including sub-Doppler lines. For certain applications, this approach has the disadvantage that suitable molecular lines are usually distant by several GHz from an arbitrary frequency. Stabilization to a ULE reference cavity in the mid-IR has not been implemented yet, to our knowledge, and would suffer from the same disadvantage. We conclude that it would be very useful to develop a tool for the MIR spectral range that provides both flexible frequency stabilization and absolute frequency measurements.

One field of research where such sources are of urgent need is trapped and sympathetically cooled molecular ions. Essentially any molecular ion species can be cooled to temperatures on the order of 10 mK by Coulomb interaction with laser-cooled atomic ions [31]. For example, for the intensely studied molecular ion $HD^+$ [4], the fundamental vibrational transition at 5 μm wavelength has a Doppler width of approx. 3 MHz. Heavier ions have a correspondingly smaller Doppler width. In the experiments on molecular ions one requires lasers able to address such transitions reliably during many hours of experimentation. This obviously implies the need for an absolute frequency instability at a level significantly less than 1 MHz, not achievable with mid-IR wavemeters. A power level of the beam to be delivered to the ions' vacuum chamber on the order of 1 mW greatly simplifies experimentation.

Precision (i.e. Doppler-free) spectroscopy on molecular ions, for example by two-photon spectroscopy, is an even more demanding application. For example, it has been proposed and analyzed in some detail for the molecular hydrogen ions [32, 33]. Here even higher levels of power, absolute frequency stability, and in addition a laser linewidth significantly below 100 kHz are necessary. The present development is motivated by such an application.

Upconversion of the MIR radiation to the NIR range in principle provides a way to take advantage of the frequency measurement capabilities of a more standard frequency comb. In previous work, the spectral range λ < 4.5 μm [34, 35, 24, 26] and λ = 10.6 μm [36, 8] were successfully covered, by using the standard nonlinear-optical material periodically poled lithium niobate and $AgGaS_2$, respectively. Very recently, a QCL emitting at 9.1 μm has been phase-stabilized to a frequency comb [30], using an approach similar to Ref. [8]. Mixing the QCL with a 2 μm Thulium comb produced an up-converted sum-frequency comb, which was beaten with an octave-spanning spectrum produced by the 2 μm comb. The beat was employed to phase-lock and linewidth-narrow the QCL, with a (1 ms) linewidth of 25 kHz.

The properties of the nonlinear-optical crystals have been a limiting factor in the flexibility and simplicity of the down-conversion and up-conversion schemes proposed so far: the limited transparency range in case of the periodically poled oxides or the difficult phase-matching conditions of other materials, requiring non-standard frequency combs.

We present a simple and robust solution applicable to the spectral range 4.5 μm < λ < 10 μm, using nearly only standard commercial components. The solution is based on the use of an advanced MIR nonlinear-optical material, orientation-patterned gallium arsenide. With the proper quasi-phasematching (QPM) period it is capable of generating the sum-frequency wave of any MIR cw laser source with a standard high-power cw C-band (NIR) laser. If the sum-frequency wave and the C-band laser are measured by a standard NIR frequency comb, it is possible to (i) determine the absolute frequency of the MIR source, (ii) determine the spectral properties of the MIR source, (iii) stabilize the frequency of the MIR source, all in real-time and with resolution and precision at the sub-kHz level. With a linewidth of 1.2 MHz, Doppler-limited spectroscopy and quantum state preparation of cold molecular ions is already possible.

## 2. Approach

Gallium arsenide (GaAs) is a suitable material for the nonlinear frequency conversion of radiation in the 1.5 – 10 μm range. Important features of the GaAs crystal are its high nonlinear-optical coefficient ($d \approx$ 100 pm/V), wide transparency range (0.9 – 17 μm in high-purity material), and high thermal conductivity (50 W·K$^{-1}$·m$^{-1}$). For lack of birefringence, one must use the QPM technique for efficient nonlinear-optical conversion. As the electric field poling technique employed for oxide crystals cannot be employed for GaAs, the crystal must be grown with the optical axis periodically reversed (orientation patterning, OP). Recently, methods were developed for fabrication of wafer-size OP-GaAs structures. They rely on a specific epitaxial growth step based on hydride-vapor-phase epitaxy (HVPE) carried out on a pre-patterned substrate [37, 38]. Cw MIR DFG sources based on OP-GaAs with output powers between nW and mW were demonstrated [39-41], and pulsed optical parametric amplification obtained with a 53 dB gain [42].

OP-GaAs crystals can also be used in the 'reversed' regime, for upconversion from the MIR to significantly shorter wavelengths, where techniques for optical detection and optical frequency metrology are well established. Thus, our approach is OP-GaAs-based frequency mixing of any MIR laser (e.g. a QCL with $\lambda_{MIR}$ in the 5 – 12 µm range) with a C-band telecom fiber laser ($\lambda_{C\text{-}band}$ = 1.54 – 1.57 µm). The upconversion of the MIR laser can be implemented in two ways:

(i) SFG ($\nu_{SFG} = \nu_{MIR} + \nu_{C\text{-}band}$), resulting in NIR upconversion wavelengths $\lambda_{SFG}$ = 1.2 – 1.4 µm, and

(ii) DFG ($\nu_{DFG} = \nu_{C\text{-}band} - \nu_{MIR}$), resulting in wavelengths $\lambda_{DFG}$ = 1.8 – 2.4 µm.

Both SFG and DFG wavelengths overlap with the spectra obtainable from commercial Er:fiber frequency comb systems.

An apparent drawback of the upconversion approach is the relatively low nonlinear conversion efficiency. However, this can be compensated by use of a C-band fiber laser having high output power. Our own work with OP-GaAs crystals shows that this material easily withstands 10 – 15 W cw laser irradiation at 1.5 µm [41]. Furthermore, the low power output of the upconversion setup can be boosted by further amplifying the upconverted wave by a semiconductor laser amplifier (at low cost) or a fiber amplifier (Pr-, Tm- doped).

The OP-GaAs crystal must be matched to the input laser wavelengths $\lambda_{MIR}$, $\lambda_{C\text{-}band}$ in order to maximize the nonlinear conversion efficiency. The coarse phase matching is achieved by proper choice of the QPM grating period of the OP-GaAs crystal. The precise phase matching can be implemented by adjustment of the crystal's temperature. It is also possible by the tuning of the C-band laser's frequency.

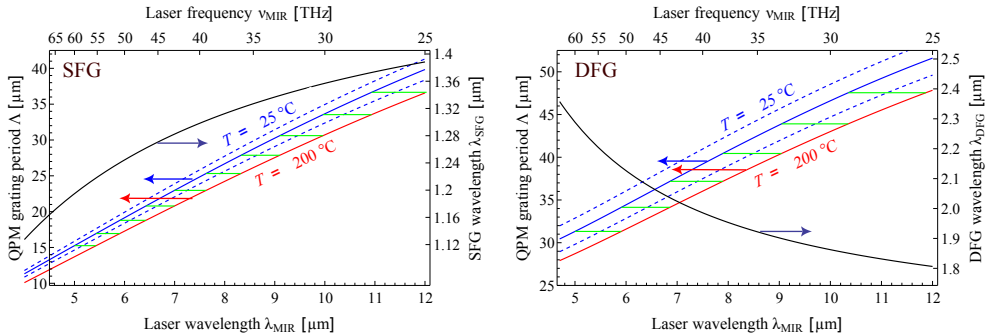

Fig. 1. First-order QPM conditions for sum frequency generation (SFG, left) or difference frequency generation (DFG, right) between a C-band laser and MIR lasers emitting in the range 5 – 12 µm.

Blue, Red (left vertical axes): 1st order QPM grating period vs. MIR laser wavelength at 25 °C and 200 °C temperature of the OP-GaAs crystal, respectively. The wavelength of the C-band laser is 1.57 µm for the solid blue and red lines. The dashed blue lines were calculated for C-band lasers at 1.54 µm and 1.60 µm, thus showing which $\lambda_{MIR}$ range can be covered by tuning of the C-band laser. Black (right vertical axes): output wavelength of the up-converted radiation $\lambda_{SFG}$ or $\lambda_{DFG}$ for a C-band laser wavelength of 1.57 µm. The green lines show the minimum number of gratings required to cover the 5 – 12 µm range by temperature tuning of the crystal. For third-order QPM, the grating period must be multiplied by 3.

The conditions for SFG and DFG frequency mixing between a C-band laser and the MIR laser are illustrated in Fig. 1. The QCL wavelength range shown in the figure is shorter than 12 µm, since this is the most important range for molecular spectroscopy. Calculations [43] were made for two temperatures of the OP-GaAs crystal: 25 °C (blue) and 200 °C (red), which represent the extremes of the temperature range at which the crystal can be operated with a standard heater.

The frequency range $\Delta\nu_{MIR}$ in which the phase-matching can be reached by the adjustment of the crystal temperature within 25 – 200 °C for a given grating period varies between 4.7 and 2.5 THz in the case of SFG (green lines in Fig. 1). Thus, the 5 – 12 µm MIR range can be covered using 10 QPM gratings in combination with temperature tuning. The number reduces to 6 for the DFG approach. In the future, multi-grating GaAs crystals containing such a moderate number of different gratings could be manufactured. In order to switch the system to a different wavelength range, the chip would then have to be shifted laterally in order to use a grating matching the target wavelength. Production of multi-grating samples with periods $\Lambda > 30$ µm is well under control today, so that DFG is possible for laser wavelengths > 5 µm in first-order QPM. SFG of laser wavelengths < 9 µm, however, requires third-order QPM, with a corresponding reduction in conversion efficiency.

## 3. Experimental apparatus

### 3.1 Optical setup

The functional elements of the apparatus are shown in Fig. 2. Physically, the setup consists of a 90 cm x 90 cm breadboard containing the QCL and SFG subunits, a 100 cm x 70 cm frequency comb breadboard, a 60 cm x 45 cm breadboard containing the optical reference, and two 19 inch racks where the comb stabilization electronics, the 1.5 µm laser system and the QCL and frequency control electronics units are located. One rack, the QCL breadboard and the frequency comb are connected by optical fibers. The system is designed to be transportable.

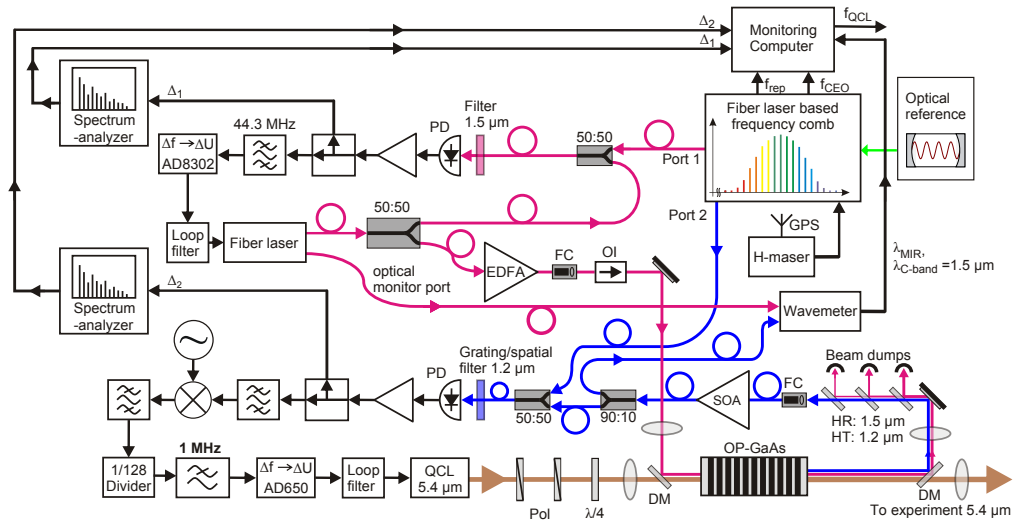

Fig. 2. Schematic of the QCL spectrometer. Colors: pink: fiber laser at 1.5 µm, brown: QCL at 5.4 µm, blue: sum frequency at 1.2 µm. DM: dichroic mirror (high transmission 5 µm, high reflectivity for 1.2 µm, 1.5 µm); SOA: semiconductor optical amplifier; Pol: wire grid polarizer; PD: photodetector; OI: optical Faraday isolator; FC: fiber collimator. The frequency comb can either be phase-locked to the optical reference for narrow linewidth or be locked to the maser only, to provide for easy tuning.

The 5.4 µm laser source is a room-temperature, distributed-feedback (DFB) QCL (Maxion) with a peak power of 40 mW, tunable from $\lambda_{QCL}$ = 1862.75 cm$^{-1}$ to 1869.25 cm$^{-1}$ at $T$ = -10 °C to 13 °C. Current is provided by a low-noise commercial current source (Wavelength Electronics, QCL1500). The laser is water-cooled and temperature-stabilized with a two-stage temperature control in the range -10 °C to 20 °C, using a home-made system. The frequency of the laser can be rapidly tuned by current modulation via a modulation input on the current source with a nominal bandwidth of 1 MHz. The output beam of the QCL is collimated by a lens attached to its front plate by the manufacturer, and its orientation can be adjusted. Unfortunately, even with an optimized lens position the beam shape is far from a Gaussian beam; it has a two-lobe structure, with about 10% in a second lobe. In order to protect the QCL from optical feedback caused by reflecting surfaces of optical elements in the setup, an optical pseudo-isolator has been set up. It consists of two wire-grid polarizers (Thorlabs) and a lambda-quarter plate (Artifex Engineering). Using only one wire-grid polarizer resulted in an isolation of more than 20 dB, which was sufficient for most optical elements like antireflection coated lenses. However, the back-reflection caused by the OP-GaAs crystal grating and its end facets still caused optical feedback in the QCL. Therefore, two wire-grid polarizers had to be used in series to suppress optical feedback sufficiently.

The C-band laser is a narrow-linewidth (ca. 3 kHz) fiber laser ($\lambda_{C\text{-}band}$ = 1565.65 nm, Koheras BasiK), with 10 mW output power, amplified to 10 W by an erbium-doped fiber amplifier (EDFA, Keopsys). The fiber laser can be tuned by means of a piezo element. The amplifier output is delivered to the breadboard via a fiber and passes an optical isolator, providing a reduced power of $P_{C\text{-}band}$ = 9 W.

The efficiency of sum-frequency generation is determined by the focusing parameters of the laser beams. The beam parameters of the two beams were measured and the best realizable focusing was determined using the formulae given in [44]. Due to astigmatism of the QCL wave, calculations for optimizing the QCL beam and 1.5 µm beam overlap were done separately for the vertical and horizontal components, each of them approximated as a Gaussian beam. Since most of the power of the QCL radiation is in one of the two lobes, the calculations were performed for the parameters determined for this lobe.

The normalized conversion efficiency $\Gamma_{SFG}$ of an SFG process is defined in terms of the input powers $P_{C\text{-}band}$, $P_{QCL}$ and the output power $P_{SFG}$ by

$$\Gamma_{SFG} = P_{SFG} / (P_{C-band} \cdot P_{QCL}).$$

$\Gamma_{SFG}$ can also be written as

$$\Gamma_{SFG} = K`_{SFG} \cdot h_{SFG}.$$

The factor $K`_{SFG}$ depends on the optical properties of the non-linear medium for each input beam, such as refractive indices, absorption coefficients and propagation length. It is also proportional to the square of the effective nonlinear-optical coefficient $d_{eff}$. The QPM requires, for our input wavelengths, a GaAs orientation patterning period of approximately 16.3 µm, which however is difficult to manufacture. Instead, we use a crystal with a period of $\Lambda$ = 49.1 µm for third-order phase matching. This results in an effective $d$-coefficient $d_{eff}$ = 2 $d$/(3 $\pi$). With the length of 17 mm, $K`_{SFG}$ = 0.0158/W results. The aperture function $h_{SFG}$ depends on the focusing parameters of the participating beams, the location of the foci within the crystal and the phase-mismatch of the conversion process. Given the beam parameters and the size constraints imposed by various optical elements, the highest possible theoretical values for $h_{SFG}$ were found to be 0.29 for the vertical and 0.26 for the horizontal components of the QCL beam.

Using a focusing lens with f = 76.6 mm, the waists in the center of the crystal are 46.5 µm and 74 µm for the vertical and horizontal beam components, respectively. The 1.5 µm radiation is focused by a 100 mm focal length lens to achieve a waist of 60 µm. The focusing lenses for both beams are mounted onto translation stages to adjust the position of the waists within the crystal. The beams are then overlapped using a dichroic mirror DM (Altechna).

The GaAs crystal was fabricated by Thales Research & Technology. It is anti-reflection coated for 1.5 µm but not for 5.4 µm and 1.2 µm. It is operated inside an oven made from copper, which is mounted on an x-y translation stage allowing adjusting the crystal perpendicularly to the beams, and a two-axis goniometer allowing adjustment of the crystal tilt relative to the beams.

The separation of the 5.4 µm beam from the 1.2 µm and 1.5 µm beams is performed directly behind the GaAs crystal by a dichroic mirror (DM). Then, around 99% of the 1.5 µm radiation is separated from the 1.2 µm radiation by another dichroic mirror (Altechna) which is highly reflective for 1.5 µm. In order to separate sufficiently well the high 1.5 µm power from the 1.2 µm beam, two more dichroic mirrors are employed to reduce the 1.5 µm power in the direction of the 1.2 µm beam to roughly 5 µW. The high-power 1.5 µm radiation is sent to beam dumps in this setup, but it could also be used for other purposes.

*3.2 Frequency stabilization*

For typical input powers in front of the crystal, $P_{C\text{-}band}$ = 9 W and $P_{QCL}$ = 25 mW, roughly one milliwatt of sum-frequency power could be expected under optimal conditions. The generated sum-frequency power is around $P_{SFG}$ = 60 µW. We believe that this is to be attributed to imperfections in the grating, the residual absorption of the OP-GaAs substrate, the non-Gaussian beam shape of the QCL, and reflection losses from the inefficient crystal coatings.

The generated sum-frequency power is not large enough for obtaining a high SNR beat note with the optical frequency comb. Therefore, a fiber-coupled semiconductor optical amplifier (SOA, Innolume GmbH) with a nominal gain of 20 dB was used to increase the sum-frequency power to 2.5 mW, which is then used to generate a heterodyne beat with the frequency comb. The output of the SOA contains a broad spontaneous emission radiation background which spans 1150 nm to 1300 nm and has a total power of about 10 mW. A substantial amount of this background radiation is removed by reflecting the heterodyne beat from a grating and partially filtering it before sending it to the photodiode.

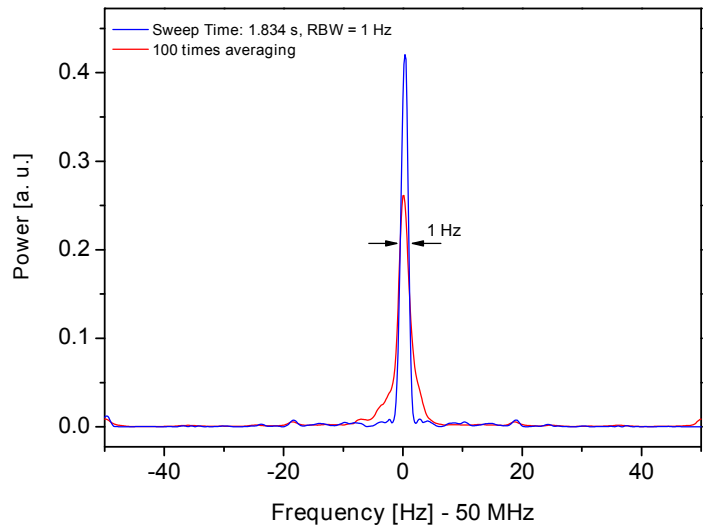

Fig. 3. Self-beat of the sum-frequency wave at 1.2 µm, between the input and the output light of the SOA. Blue: 1.8 s integration time, red: 100 averages (183 s).

In order to verify that the SOA does not add any frequency noise to the SFG wave, the amplified 1.2 µm SHG light was frequency-shifted by an AOM by 50 MHz and heterodyned in a fiber splitter/combiner with a small part of the light before the amplifier. The FWHM linewidth of this self-beat is shown in Fig. 3 and was 1 Hz, limited by the lowest bandwidth available from the spectrum analyzer (Agilent E4440A). Even at 183 s averaging time the linewidth remained at 1 Hz. This verifies that the frequency noise properties of the SOA do not add frequency noise to the sum frequency wave.

The frequency comb is based on a 250-MHz erbium-doped fiber oscillator (FC1500-250-WG, Menlo Systems). The comb is completed by three EDFA modules and a supercontinuum generation module for spectral broadening to the near-infrared (NIR) 1050 - 2100 nm spectral range. One EDFA is used as part of the carrier offset frequency ($f_{ceo}$) stabilization, one for two NIR output ports with combined average output power >200 mW and one for an additional port which is not used here. The last two generate a high power of >5 mW in a 3 nm wavelength window, allowing optional phase-locking of the comb to an optical reference. The long-term stable optical reference has 1-Hertz linewidth and consists of a Nd:YAG laser which is stabilized to an ultra-high-finesse ULE cavity.

Depending on the usage scenario, the comb operating mode can be selected. For investigation of wide (GHz) transitions, the comb repetition rate ($f_{rep}$) and the carrier offset frequency ($f_{ceo}$) are actively stabilized to a 10 MHz rf-signal provided by an ultrastable hydrogen maser which itself is steered to GPS. The repetition rate can be set and controlled from the comb's computer. This provides the tuning of the comb. The comb mode frequencies are given by $f_{ceo} + m\,f_{rep}$, with integer $m$.

If instead a narrow linewidth of the comb teeth is required, the comb can be phase-locked to an optical reference, but then it cannot be easily tuned by changing the repetition rate. The MIR- laser is then tuned by changing the beat frequency.

The amplified up-converted 1.2 μm wave and the 1.5 μm wave of the fiber laser are heterodyned in two beat note units with the respective nearest frequency comb modes. In each unit (Fig. 2), an output of the comb and a laser wave are input via respective fibers and combined in a single-mode 50% - 50% fiber splitter/combiner, allowing maximum spatial overlap. The output of the up-converted wave part is delivered into a free beam after the fiber combiner. Mode cleaning of the comb spectrum and wavelength filtering of the SOA radiation background is done by a grating. A λ/2 plate matches the polarization to the grating. In case of the fiber laser the comb spectrum is filtered due to a narrowband band-pass filter before coupling into the fiber combiner. The polarization of the comb spectrum is matched to the lasers' polarization by λ/2 plates also before coupling into the fiber combiners. Both heterodyne beats are detected by high-speed photodetectors. The beat notes are electrically amplified to an absolute signal level of 0 dBm. Subsequent directional couplers deliver the two beat notes $\Delta_1$, $\Delta_2$ to two spectrum analyzers for readout and data logging.

The beat note of the fiber laser with the comb easily reaches an SNR of 40 dB because of the small linewidth and the high power (several milliwatts) of the fiber laser beam. The beat notes are $\Delta_1 = \nu_{C\text{-}band} - (f_{ceo} + m_1\,f_{rep})$ in case of the fiber laser and $\Delta_2 = \nu_{SFG} - (f_{ceo} + m_2\,f_{rep})$ for the upconverted wave, respectively, where $m_1$ and $m_2$ are the mode numbers of the closest comb modes. The laser frequency stabilization loops keep $\Delta_1$ and $\Delta_2$ constant in time (see below).

The frequency-to-voltage detector for error signal generation in the fiber laser stabilization loop employs the Analog Devices AD8302 RF/IF gain and phase detector [45] and has a fixed locking point of $\Delta_1$ = 44.3 MHz. For a given comb repetition rate, this frequency is initially set by fine frequency tuning of the fiber laser. An active band-pass filter before the detector filters the beat note so as to lower the noise floor.

The beat note of the upconverted 1.2 μm wave, $\Delta_2$ is kept at 40 MHz by the QCL frequency lock. For a given comb repetition rate, this frequency is also initially set by fine frequency tuning of the QCL. For the stabilization loop it is further mixed to 64 MHz. This beat note SNR is limited to 20 - 25 dB due to the comparatively high linewidth of 1.2 MHz and the lower absolute power.

Appropriate bandpasses before and after the frequency mixer filter and isolate the beat signal, which is subsequently divided by 128 in a frequency divider, to a frequency of 500 kHz. The division of the beat note signal frequency enables a wide capturing range and a robust locking even with the low SNR of the beat note. The frequency-to-voltage detector in the QCL locking loop is based on the Analog Devices AD650 converter and has a detection range from 0 to 1 MHz, with a 0 V output for a 500 kHz signal input.

The frequencies of the QCL and of the fiber laser are each stabilized by two frequency-locking systems, adapted to the differing characteristics of the respective beat notes.

For the stabilization of the QCL, the modulation input of the current driver is used. The standard PI servo for the fiber laser acts on the piezo element inside the laser. For fixed frequency operation, the piezo control voltage stays within the allowed working range. In order to be able to scan the frequency of the fiber laser, a LabView program is used to keep the piezo control voltage within the piezo's specs by tuning the temperature setpoint of the fiber laser via its internal microcontroller.

Since long-term drifts in the beat signal frequency of the sum frequency light and the frequency comb (but not in the 1.5 µm beat) were observed during initial long-term stability measurements, a LabView-based PI control was added that adjusts the frequency of the synthesizer used to mix down the beat signal. This ensures that the beat frequency stays constant and also keeps the QCL frequency constant.

*3.3 Frequency measurement and tuning*

The optical frequency of the QCL is determined by the relation

$$\nu_{QCL} = (m_2 - m_1) f_{rep} \pm \Delta_2 \pm \Delta_1. \qquad (1)$$

As can be seen, the carrier envelope offset frequency $f_{ceo}$ drops out. Since all quantities on the right hand side are constant or actively kept constant in time, the QCL frequency is constant, too. The mode number difference $m_2 - m_1$, which remains constant during operation, must be known as well. $m_1$ and $m_2$ and the signs of the beat frequencies are independently determined from independent measurements of the wavelengths $\lambda_{SFG}$ and $\lambda_{C\text{-}band}$ using a wavelength meter (Burleigh WA-1500) having an inaccuracy of less than 100 MHz. A monitor fiber output port of the fiber laser and 90:10 fiber splitter behind the 1.2 µm amplifier provide radiation of sufficient power for the wavemeter. The wavemeter is also read out by the monitoring computer and the QCL frequency is determined in the computer by applying Eq. (1).

Since $f_{rep}$ is phase-locked to the atomic reference signal, which is also a reference for the spectrum analyzers that measure $\Delta_1$ and $\Delta_2$, the QCL frequency is determined in units of the atomic reference frequency, which here is 10 MHz.

The beat frequencies $\Delta_1$, $\Delta_2$ are measured every second by determining the weighted mean frequency of the beat spectra acquired by the two spectrum analyzers, set to average over 0.1 - 1 s, depending on the speed and step size of the repetition rate tuning, if any. This is an adequate alternative approach to the use of a frequency counter, in view of the fact that the 1.2 µm beat spectrum is rather wide. Thus, the QCL frequency is determined in real time.

Under lock, a change of the frequency comb repetition rate $f_{rep}$ shifts the QCL frequency with a tuning coefficient given by $m_2 - m_1$. Thus, the QCL frequency can be electronically tuned by changing the frequency comb repetition rate $f_{rep}$. This is done by the same computer which records the two beat frequencies.

## 4. Results

The phase-matching temperature for the concrete SFG process used here was found to be 44 °C. However, it varies by about 2 °C depending on the power of the 1.5 µm laser, owing to absorption in the crystal at this wavelength.

8 / 14

The beat notes with the frequency comb when stabilized to an optical reference are shown in Fig. 4. The frequency-locked fiber laser has a linewidth of 35 kHz. The beat of the amplified 1.2 μm wave with the comb essentially represents the frequency spectrum of the QCL itself, since the linewidths of the fiber laser is relatively small. The beat linewidth is comparable to the free-running linewidth that we observed over a timescale of hundreds of ms, since the bandwidth of the employed QCL locking system is relatively low and not suitable for linewidth narrowing.

The excellent long-term stability of the comb-stabilized QCL is shown in Fig. 5. Here, the comb was stabilized to the maser only. The QCL was not tuned and the beat frequencies and the comb repetition rate value were measured once per second. Thus, the QCL frequency is computed once per second. The Allan deviation, a measure of frequency instability, is less than 10 kHz for integration times longer than 40 s. The inaccuracy of the QCL frequency is estimated as 100 kHz.

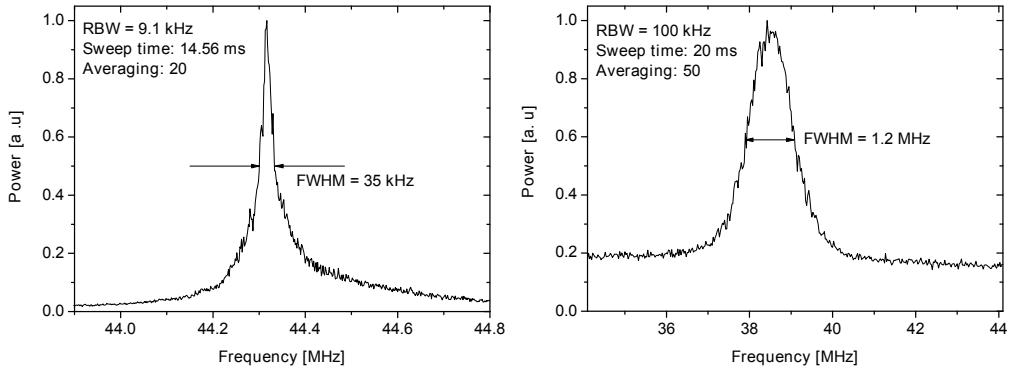

Fig. 4. Beat notes with the frequency comb which is stabilized to an optical reference. Left: Beat note of the 1.5 μm fiber laser and the frequency comb line to which it is weakly locked. Right: Beat note between the sum-frequency wave at 1.2 μm and a frequency comb line, for 1 s averaging time.

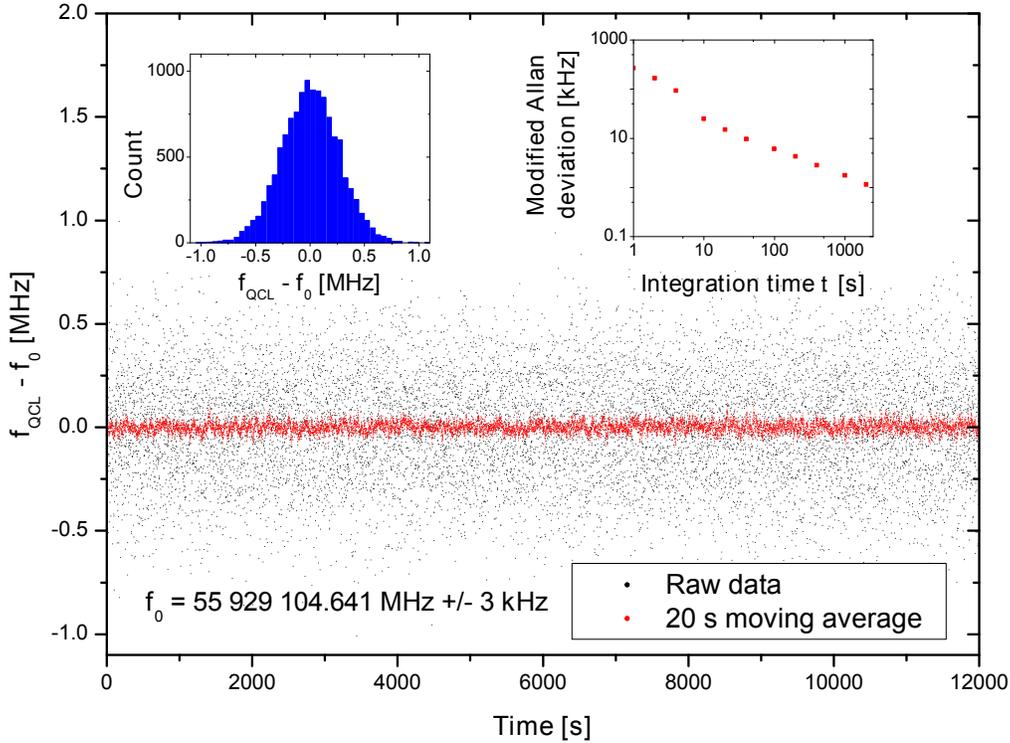

Fig. 5. Stability of the QCL frequency over three hours and histogram of frequency values and Allan deviation (insets).

The usability of the developed source for spectroscopic measurements is demonstrated by the results in Fig. 6. The QCL frequency is scanned via computer-controlled change of the comb repetition rate. This method provides a wide tuning range of at least 2.5 GHz. On purpose, we scanned the frequency very slowly, in 5000 s, to show the capability of long measurement times. The panel at the bottom of Fig. 6 shows the frequency deviations from a linear variation in time; the deviations are similar to those reported in Fig. 5 in absence of frequency scan.

In the test experiment, the QCL interrogates the P17E absorption line of the (1 1 1 0) ← (0 0 0 0) band of $N_2O$ in a gas cell at 60 mbar. Before the QCL beam is sent through the cell, a part of the beam is split off the main beam and sent to a detector for normalization. The absorption line is shown in Fig. 6. Since the baseline of the data is not constant, a Voigt profile fit is done only using the central 600 MHz interval around the absorption peak. The fitted line center is at 55 929 343.8 ± 0.3 MHz, where the uncertainty is the 1-σ range given by the standard error of the fit (one scan). The NIST wavenumber calibration tables give 55 929 347.6 ± 2.5 MHz at a pressure of 2.7 mbar [46]). The pressure shift of the line at our pressure can be estimated to be in the range of -0.3 MHz to -5.3 MHz from pressure shifts given for other lines of the same band [47]. The difference between the two measurements therefore lies in the range of -3.4 MHz- to +1.5 Mhz, and is consistent with zero within the error of the NIST value. However, we emphasize that our experiment was not designed for high accuracy, but only for demonstration.

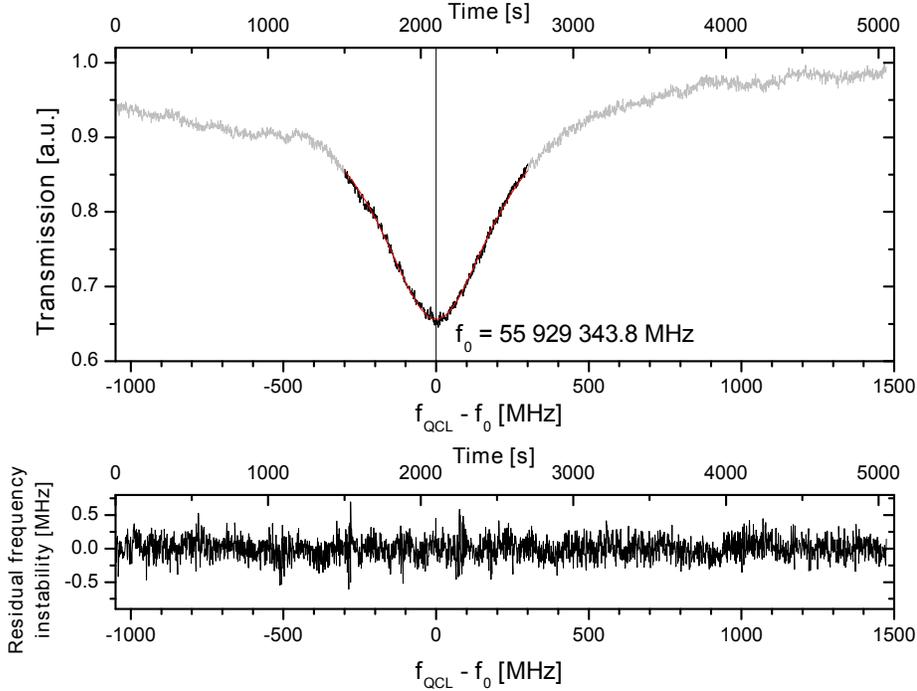

Fig. 6. Spectrometer frequency scan over a range of 2.5 GHz at a wavelength of 5.36 μm, performed by scanning the repetition rate of the frequency comb. An absorption line of $N_2O$ was recorded simultaneously. Top: Grey line: Transmission through the gas cell versus frequency. A Voigt profile (red line) is fit to the central part of the peak (black part of the absorption peak). Bottom: measured QCL frequency minus linear fit of the frequency vs. time, showing the residual instability during the frequency scan.

## 6. Summary

In this paper, we have demonstrated a general method capable of providing frequency-stable radiation in the MIR range from quantum cascade lasers, with frequency absolutely known at the 100 kHz level and stable at the few kHz level. The absolute frequency is measured relative to an atomic reference.

Key element is an orientation-patterned-GaAs crystal for up-conversion of the QCL laser radiation. The crystal's properties were matched to the laser wavelength $\lambda_{QCL}$ to be measured, in order to maximize the conversion efficiency of the SFG process. The coarse phase matching of the SFG process was achieved by proper choice of the QPM grating period of the OP-GaAs crystal. The precise matching was implemented by adjustment of the crystal's temperature. The fine tuning of the phase matching is also possible by tuning of the local oscillator laser's frequency.

The generated SFG radiation at 1.2 μm was amplified by using a semiconductor amplifier to a level of ca. 2 mW. We have shown that the amplification does not degrade the spectral properties. A beat with a phase-stabilized Erbium fiber frequency comb was thereby obtained with sufficient signal-to-noise-ratio. The beat permits measuring and correcting the absolute frequency fluctuations of the QCL with (currently) few kHz-level resolution.

While we have demonstrated the method for a particular MIR wavelength, and our crystal only contained two gratings, we have shown theoretically that a moderate number of OP-GaAs gratings are sufficient to cover the complete 5 – 12 μm spectral range.

The implementation of the method is robust and relatively simple. All components except the OP-GaAs crystal are standard and robust, i.e. frequency comb, atomic reference, 1.5 µm cw high-power single-frequency fiber laser, semiconductor amplifier, detectors, etc.. The whole system requires only short warm-up time, minimal realignments in day-to-day use and is locked within a few minutes. This is an important advantage for use of the apparatus as part of more complex experimental set-ups.

The full power of the QCL is sent through the nonlinear crystal, but, apart from reflection losses at interfaces, most of it is recovered after the crystal, and can be delivered to a spectroscopy apparatus. This means that sufficient power remains available to implement, e.g. saturation or photoacoustic spectroscopic techniques, which have higher power requirements than standard absorption spectroscopy.

In conclusion, we believe that the presented approach represents an important and flexible tool for MIR precision (i.e. comb-assisted) spectroscopy.

With its current performance and its ease of use, this type of spectrometer could be used for photoacoustic spectroscopy, multipass-cell spectroscopy, integrated cavity output spectroscopy, or Lamb-dip spectroscopy. But even as part of more sophisticated experiments, for example in the field of cold molecules [31], where the low particle temperatures require MHz-level laser linewidths, this system is suitable.

In the near future, after the development of a suitable feedback system capable of reducing further the linewidth of the QCL to the kHz level and below, high-resolution studies of molecular transitions will become possible with methods such as cavity-ring-down spectroscopy, NICE-OHMS [48], where efficient coupling into high-finesse (narrow-linewidth) optical resonators is necessary, or two-photon spectroscopy, where ultra-narrow linewidth and high power are necessary.

### Acknowledgments

We thank T. Schneider and P. Dupré for helpful suggestions on the setup, and D. Iwaschko for development of electronic units. The optical reference cavity has been developed by A. Nevsky, Q. Chen, M. Cardace (Universität Düsseldorf) and U. Sterr (Physikalisch-Technische Bundesanstalt) as part of project 50OY1201 funded by the Bundesministerium für Wirtschaft und Technologie (Germany). This work was partially funded by DFG project SCHI 431/19-1.